\documentclass{article}
\usepackage{graphicx}
\usepackage{amsmath}

\input{tcilatex}

\begin{document}

\title{SOLUTIONS WITHOUT SINGULARITIES IN GAUGE THEORY OF GRAVITATION}
\author{G. ZET$^{\ast }$ and C.\ D. OPRISAN \\
Department of Physics,\\
''Gh. Asachi'' Technical University,\\
Iasi 6600, Romania\\
$^{\ast }$gzet@phys.tuiasi.ro\\
S. BABETI\\
Department of Physics,\\
''Politehnica'' University, \\
Timisoara 1900, Romania}
\maketitle

\begin{abstract}
A de-Sitter gauge theory of the gravitational field is developed using a
spherical symmetric Minkowski space-time as base manifold. The gravitational
field is described by gauge potentials and the mathematical structure of the
underlying space-time is not affected by physical events. The field
equations are written and their solutions without singularities are obtained
by imposing some constraints on the invariants of the model. An example of
such a solution is given and its dependence on the cosmological constant is
studied. A comparison with results obtained in General Relativity theory is
also presented.

Keywords: gauge theory, gravitation, singularity, computer algebra
\end{abstract}

\section{Introduction}

The gauge theory of gravitation has been considered by many authors in order
to describe the gravity in a similar way with other interactions
(electromagnetic, weak or strong). As gauge groups there were chosen
Poincar\'{e} group$^{1}$, de-Sitter group$^{2}$, affine group$^{3}$, etc. In
this paper we use the de-Sitter ($DS$) group as gauge group in order to
obtain a model with cosmological constant for the gravitational field. The
Poincar\'{e} gauge theory is obtained as a limit of $DS$ model when the
cosmological constant vanishes.

In Sect.1 we introduce the gauge fields $e_{\mu }^{a}\left( x\right) $
(tetrad fields) and $\omega _{\mu }^{ab}\left( x\right) $ (spin connection).
They are used to construct the field strengths $F_{\mu \nu }^{a}$ and $%
F_{\mu \nu }^{ab}$ and the invariants of the theory. Then the integral of
the action is written and the constraints for non-singular solutions are
introduced in its expression by means of two Lagrange-multiplier fields $%
\varphi _{1}\left( t\right) $ and $\varphi _{1}\left( t\right) $.

The field equations are obtained in Sect.2 for a particular form of
spherically symmetric gauge fields. They contain the cosmological constant $%
\Lambda $ introduced into the model by using the $DS$ group as gauge
symmetry. The calculations in this paper, especially in Sect. 3, have been
performed using an analytical program written by us in the package GRTensor
II running on the Maple V platform. This program allows to calculate the
components of the field strengths, the invariants of the model and also the
integral of the action. In the same time, it enables the obtaining of the
field equations for the gravitational field in a region without matter.

An example of solution without singularities is presented in Sect. 3. This
solution is a time-periodic one with frequency of the gravitational field
depending on the cosmological constant. It corresponds to a negative value
of the cosmological constant ($\Lambda <0$). The case with positive
cosmological constant ($\Lambda >0$) can be studied choosing the
anti-de-Sitter group as gauge group.

In Sect.4 some concluding remarks are presented and a comparison with other
results based on the General Relativity theory is made.

\section{Gauge theory of gravitation}

We consider a gauge theory of gravitation having de-Sitter ($DS$) group as
local symmetry. Let $X_{A}$, $A=1,2,....,10,$ be a basis of $DS$ Lie algebra
with the corresponding equations of structure given by$^{2}$: 
\begin{equation}
\left[ X_{A},X_{B}\right] =if_{AB}^{C}X_{C},  \tag{1}
\end{equation}
where $f_{AB}^{C}=-f_{BA}^{C}$ are the constants of structures whose
expressions will be given \ below [see Eq.(4)]. We envision space-time as a
four-dimensional manifold $M_{4}$; at each point of $M_{4}$ we have a copy
of $DS$ group (i.e., a fibre, in fibre-bundle terminology). Introduce, as
usually, the gauge potentials $h_{\mu }^{A}\left( x\right) $, $%
A=1,2,....,10,\mu =0,1,2,3$, were $\left( x\right) $ denotes the local
coordinates on $M_{4}$. Then, we calculate the field-strengths $F_{\mu \nu
}=F_{\mu \nu }^{A}X_{A}$, which take values in Lie algebra of $DS$ group
(Lie-algebra valued). The components $F_{\mu \nu }^{A}$ are given by: 
\begin{equation}
F_{\mu \nu }^{A}=\partial _{\mu }h_{\nu }^{A}-\partial _{\nu }h_{\mu
}^{A}+f_{BC}^{A}h_{\mu }^{B}h_{\nu }^{C}\text{ .}  \tag{2}
\end{equation}

In order to write the constants of structure $f_{AB}^{C}$ in a compact form,
we use the following notations for the index $A$: 
\begin{equation}
A=\left\{ 
\begin{array}{c}
a=0,1,2,\text{\ \ \ \ \ \ \ \ \ \ \ \ \ \ \ \ \ \ \ \ \ \ \ \ } \\ 
\lbrack ab]=[01],[02],[03],[12],[13],[23].
\end{array}
\right.  \tag{3}
\end{equation}
This means that $A$ can stand for a single index like 2 as well as for a
pair of indices like $\left[ 01\right] $, $\left[ 12\right] $ etc. The
infinitesimal generators $X_{A}$ are interpreted as: $X_{a}\equiv P_{a}$
(energy-momentum operators) and $X_{[ab]}\equiv M_{ab}$ (angular momentum
operators) with the property $M_{ab}=-M_{ba}$. The constants of structures $%
f_{AB}^{C}$ have then the following expressions: 
\begin{eqnarray}
f_{bc}^{a} &=&f_{c[de]}^{[ab]}=f_{[bc][de]}^{a}=0,  \TCItag{4} \\
f_{cd}^{[ab]} &=&4\lambda ^{2}\left( \delta _{c}^{b}\delta _{d}^{a}-\delta
_{c}^{a}\delta _{d}^{b}\right) =-f_{dc}^{[ab]},  \notag \\
f_{b[cd]}^{a} &=&-f_{[cd]b}^{a}=\frac{1}{2}\left( \eta _{bc}\delta
_{d}^{a}-\eta _{bd}\delta _{c}^{a}\right) ,  \notag \\
f_{[ab][cd]}^{[ef]} &=&\frac{1}{4}\left( \eta _{bc}\delta _{a}^{e}\delta
_{d}^{f}-\eta _{ac}\delta _{b}^{e}\delta _{d}^{f}+\eta _{ad}\delta
_{b}^{e}\delta _{c}^{f}-\eta _{bd}\delta _{a}^{e}\delta _{c}^{f}\right)
-e\longleftrightarrow f,  \notag
\end{eqnarray}
where $\lambda $ is a real parameter, and ($\eta _{ab})=$diag$(1,-1,-1,-1)$
is the Minkowski metric. In fact, here we have a deformation of de-Sitter
Lie algebra having $\lambda $ as parameter. Considering the contraction $%
\lambda \longrightarrow 0$ we obtain the Poincar\'{e} Lie algebra, i.e., the
group DS contracts to the Poincar\'{e} group.

We denote the gravitational gauge fields (or potentials), $h_{\mu
}^{A}\left( x\right) $, by $e_{\mu }^{a}(x)$ (tetrad fields) if $A=a$, and
by $\omega _{\mu }^{ab}(x)=-\omega _{\mu }^{ba}(x)$ (spin connection) if $%
A=[ab]$. Then, introducing the Eqs. (4) into the definition (2), we obtain
the expressions of the strength tensor components$^{2}$: 
\begin{equation}
F_{\mu \nu }^{a}=\partial _{\mu }e_{\nu }^{a}-\partial _{\nu }e_{\mu
}^{a}+\left( \omega _{\mu }^{ab}e_{\nu }^{c}-\omega _{\nu }^{ab}e_{\mu
}^{c}\right) \eta _{bc}  \tag{5}
\end{equation}
\begin{equation}
F_{\mu \nu }^{ab}=\partial _{\mu }\omega _{\nu }^{ab}-\partial _{\nu }\omega
_{\mu }^{ab}+\left( \omega _{\mu }^{ac}\omega _{\nu }^{db}-\omega _{\nu
}^{ac}\omega _{\mu }^{db}\right) \eta _{cd}-4\lambda ^{2}\left( e_{\mu
}^{a}e_{\nu }^{b}-e_{\nu }^{a}e_{\mu }^{b}\right) .  \tag{6}
\end{equation}
The integral of action associated to the gravitational gauge fields,
quadratic in the components $F_{\mu \nu }^{A}$, is written in the form$^{2}$%
: 
\begin{equation}
S_{g}=\int d^{4}x\varepsilon ^{\mu \nu \rho \sigma }F_{\mu \nu }^{A}F_{\rho
\sigma }^{B}Q_{AB},  \tag{7}
\end{equation}
where $\varepsilon ^{\mu \nu \rho \sigma }$ is the Levi-Civita symbol of
rank four, with $\varepsilon ^{0123}=1$. This action is independent of any
specific metric on $M_{4}$; indeed, the property of general covariance for
action imposes the volume element $\sqrt{-g}d^{4}x$ [with $g=\det (g_{\mu
\nu })$] and the tensor Levi-Civita has the form $\varepsilon ^{\mu \nu \rho
\sigma }/\sqrt{-g}$, so that the $g_{\mu \nu }$-dependence of $S_{g}$
cancels.

The quantities $Q_{AB}$ are constants, symmetric with respect to the indices 
$A,B$: $Q_{AB}=$ $Q_{BA}$. If we chose$^{4}$%
\begin{equation}
Q_{AB}=\left\{ 
\begin{array}{c}
\varepsilon _{abcd},\text{ for \ \ }A=\left[ ab\right] ,\text{ \ }B=\left[ cd%
\right] , \\ 
0\text{ \ \ \ \ \ \ \ \ \ \ \ \ \ \ \ \ \ \ \ \ \ \ \ \ otherwise},
\end{array}
\right.  \tag{8}
\end{equation}
then we obtain the action integral of the General Relativity ($GR$). It is
possible also to obtain the integral action of Teleparallel Gravity ($TG$)
by an appropriate choice$^{5,6}$ of $Q_{AB}$.

Now, we use the form given in Eq. (8) in order to obtain solutions without
singularities of $DS$-gauge theory of gravitation. Namely, we impose some
restrictions$^{7}$ on two invariants $I_{1}$ and $I_{2}$ of the theory.
Introducing the Lagrange-multiplier $\varphi _{1}\left( t\right) $ and $%
\varphi _{2}\left( t\right) $, and using the choice (8), the integral of
action (7) can be rewritten as: 
\begin{equation}
S_{g}=-\frac{1}{16\pi G}\int d^{4}xe\left[ F+\varphi _{1}\left( t\right)
f_{1}\left( I_{1}\right) +\varphi _{2}\left( t\right) f_{2}\left(
I_{2}\right) +V\left( \varphi _{1},\varphi _{2}\right) \right] ,  \tag{9}
\end{equation}
where 
\begin{equation}
F=F_{\mu \nu }^{ab}\overline{e}_{a}^{\mu }\overline{e}_{b}^{\nu },\text{ \ \ 
}e=\det \left( e_{\mu }^{a}\right) .  \tag{10}
\end{equation}
and $\overline{e}_{b}^{\nu }$ is the inverse of $\ e_{\mu }^{a}$ defined by
Eq. (21) below. The quantities $f_{i}\left( I_{i}\right) $, $i=1,2$ are
functions which must be chosen in an appropriate form in order to obtain
solutions without singularities of the corresponding field equations. Thus,
the potential $V\left( \varphi _{1},\varphi _{2}\right) $ have to satisfy
the constraint equations$^{7}$: 
\begin{equation}
f_{1}\left( I_{1}\right) =-\frac{\partial V}{\partial \varphi _{1}},\text{ \
\ }f_{2}\left( I_{2}\right) =-\frac{\partial V}{\partial \varphi _{2}}, 
\tag{11}
\end{equation}
The model can be simplified further if we assume: 
\begin{equation}
V\left( \varphi _{1},\varphi _{2}\right) =V_{1}\left( \varphi _{1}\right)
+V_{2}\left( \varphi _{2}\right) ,  \tag{12}
\end{equation}
and chose the invariants $I_{1}$, $I_{2}$ in the form 
\begin{equation}
I_{1}=F-\sqrt{3}\left( 4F_{\mu }^{a}F_{a}^{\mu }-F^{2}\right) ^{1/2}, 
\tag{13}
\end{equation}
respectively 
\begin{equation}
I_{2}=4F_{\mu }^{a}F_{a}^{\mu }-F^{2}.  \tag{14}
\end{equation}
In these expressions, the quantities $F_{\mu }^{a}$ are defined by 
\begin{equation}
F_{\mu }^{a}=F_{\mu \nu }^{ab}\overline{e}_{b}^{\nu }.  \tag{15}
\end{equation}
\qquad

As an example, we chose the functions $f_{1}$ and $f_{2}$ in the simple form$%
^{7}$: 
\begin{equation}
f_{1}\left( I_{1}\right) =I_{1},\text{ \ \ }f_{2}\left( I_{2}\right) =-\sqrt{%
I_{2}}.  \tag{16}
\end{equation}
Then, the action $S_{g}$ in Eq. (9) becomes: 
\begin{equation}
S_{g}=-\frac{1}{16\pi G}\int d^{4}xe\left[ F+\varphi _{1}I_{1}-\varphi _{2}%
\sqrt{I_{2}}+V_{1}\left( \varphi _{1})+V_{2}(\varphi _{2}\right) \right] . 
\tag{17}
\end{equation}

Now, all we have to do is to write the variational field equations which
follow from (17) and search their solutions without singularities.

\section{Field equations}

We develop the $DS$ gauge theory in a space-time Minkowski $M_{4}$ endowed
with spherical symmetry: 
\begin{equation}
ds^{2}=dt^{2}-dr^{2}-r^{2}\left( d\theta ^{2}+\sin ^{2}\theta d\varphi
^{2}\right)   \tag{18}
\end{equation}
and having the coordinates $\left( x^{\mu }\right) =\left(
x^{0},x^{1},x^{2},x^{3}\right) =\left( t,r,\theta ,\varphi \right) $. In
addition, we choose a particular form of spherically gauge fields $e_{\mu
}^{a}\left( x\right) $ and $\omega _{\mu }^{ab}\left( x\right) $ given by
the following ansatz: 
\begin{equation}
e_{\mu }^{0}=\left( N(t),0,0,0\right) ,\text{ \ \ }e_{\mu }^{1}=\left( 0,%
\frac{a\left( t\right) }{\sqrt{1-kr^{2}}},0,0\right) ,\text{ \ }  \tag{19a}
\end{equation}
\begin{equation}
\text{\ }e_{\mu }^{2}=\left( 0,0,ra\left( t\right) ,0\right) ,\text{ \ \ \ }%
e_{\mu }^{2}=\left( 0,0,0,ra\left( t\right) \sin \theta \right) ,  \tag{19b}
\end{equation}
respectively 
\begin{equation}
\omega _{\mu }^{01}=\left( 0,-\frac{a^{\prime }\left( t\right) }{N\left(
t\right) \sqrt{1-kr^{2}}},0,0\right) ,\text{ \ \ }\omega _{\mu }^{02}=\left(
0,0,-\frac{ra^{\prime }\left( t\right) }{N\left( t\right) },0\right) , 
\tag{20a}
\end{equation}
\begin{equation}
\text{\ }\omega _{\mu }^{03}=\left( 0,0,0,-\frac{ra^{\prime }\left( t\right)
\sin \theta }{N\left( t\right) }\right) ,\text{ \ \ \ \ }\omega _{\mu
}^{12}=\left( 0,0,\sqrt{1-kr^{2}},0\right) ,  \tag{20b}
\end{equation}
\begin{equation}
\omega _{\mu }^{13}=\left( 0,0,0,\sqrt{1-kr^{2}}\sin \theta \right) ,\text{
\ \ }\omega _{\mu }^{23}=\left( 0,0,0,\cos \theta \right)   \tag{20c}
\end{equation}
where $N\left( t\right) $ and $a\left( t\right) $ are functions only of the
time variable, $k$ is a constant, and $a^{\prime }$ is the derivative of \ $%
a\left( t\right) $ with respect to the variable $t$. The choice (20) of
gauge fields $\omega _{\mu }^{ab}\left( x\right) $ assures that all
components of the strength tensor $F_{\mu \nu }^{a}$ vanish. If we remember
the Riemann-Cartan theory of gravitation, then this result implies the
vanishing of the torsion tensor $T_{\mu \nu }^{\rho }=\overset{\_}{e}%
_{a}^{\rho }F_{\mu \nu }^{a}$, in accord with $GR$ theory. Here, $\overset{\_%
}{e}_{a}^{\rho }$denotes the inverse of $e_{\mu }^{a}$ with the properties: 
\begin{equation}
e_{\mu }^{a}\overset{\_}{e}_{b}^{\mu }=\delta _{b}^{a},\text{ \ \ }e_{\mu
}^{a}\overset{\_}{e}_{a}^{\nu }=\delta _{\mu }^{\nu }.  \tag{21}
\end{equation}

From this point to the end we performed all the calculations using an
analytical program conceived by us which is presented in the final part of
this Section.

Using the Eqs. (19) and (20), we obtain the following expressions of the
invariants $F$, $I_{1}$ and $I_{2}$ above defined: 
\begin{equation}
F=-6\frac{aa^{\prime \prime }N-aa^{\prime }N^{\prime }+kN^{3}+a^{\prime
2}N+8\lambda ^{2}a^{2}N^{3}}{a^{2}N^{3}},  \tag{22}
\end{equation}
\begin{equation}
I_{1}=-12\frac{kN^{2}+a^{\prime 2}+4\lambda ^{2}a^{2}N^{2}}{a^{2}N^{2}}, 
\tag{23}
\end{equation}
and respectively 
\begin{equation}
I_{2}=12\frac{\left( kN^{3}+a^{\prime 2}N-aa^{\prime \prime }N+aa^{\prime
}N^{\prime }\right) ^{2}}{a^{4}N^{6}}.  \tag{24}
\end{equation}
where $a^{\prime \prime }$ is the second derivative of $a\left( t\right) $
with respect to the variable $t$. Introducing these expressions into Eq.
(17) and imposing the variational principle $\delta S_{g}=0$ with respect to 
$N\left( t\right) $, $\varphi _{1}\left( t\right) $ and $\varphi _{2}\left(
t\right) $, we obtain the corresponding field equations. We write now these
equations for the particular case $N\left( t\right) =1$ which is of interest
in our model: 
\begin{equation}
\left\{ 
\begin{array}{c}
-\frac{1}{2}\left( V_{1}+V_{2}\right) +3H^{2}\left( 1-2\varphi _{1}\right) +3%
\frac{k}{a^{2}}\left( 1+2\varphi _{1}\right) -2\Lambda = \\ 
\sqrt{3}\left( \varphi _{2}^{\prime }+3H\varphi _{2}-\frac{k}{Ha^{2}}\varphi
_{2}\right) ,
\end{array}
\right.   \tag{25}
\end{equation}
\begin{equation}
\frac{k}{a^{2}}+H^{2}-\frac{\Lambda }{3}=\frac{1}{12}\frac{dV_{1}}{d\varphi
_{1}},\text{ \ \ }H=\frac{a^{\prime }}{a},  \tag{26}
\end{equation}
\begin{equation}
H^{\prime }-\frac{k}{a^{2}}=-\frac{1}{2\sqrt{3}}\frac{dV_{2}}{d\varphi _{2}},%
\text{ \ \ }H^{\prime }=\frac{dH}{dt}=\frac{a^{\prime \prime }a-a^{\prime 2}%
}{a^{2}},  \tag{27}
\end{equation}
where $\varphi _{2}^{\prime }$ is the derivative of $\varphi _{2}\left(
t\right) $ with respect to $t$, and $\Lambda =-12\lambda ^{2}$ is
interpreted as cosmological constant.$^{2,8}$

If we consider the limit $\lambda \longrightarrow 0$, or equivalently $%
\Lambda =0$, we obtain the results in Ref.$\left[ 7\right] $; but, for $%
\Lambda \neq 0$ we can study in addition the dependence on the cosmological
constant of the solutions (without singularities) obtained by solving the
Eqs.(25)-(27). We make also the mention that the Eqs. (26) and (27) are
identically with the constraints (11) introduced into the integral of action 
$S_{g}$ by means of the Lagrange-multiplier fields $\varphi _{1}\left(
t\right) $ and $\varphi _{2}\left( t\right) $.

We can also add matter to the previous model considering the integral of
action: 
\begin{equation}
S_{m}=\int d^{4}xeL_{m},  \tag{28}
\end{equation}
where $L_{m}$ is the matter density of Lagrangian. In this paper we restrict
ourselves to the case without matter. In Section 4 we will obtain a
particular solution with fixed cosmological \ constant $\Lambda =$const. Of
course, there are possible also solutions with variable cosmological
''constant'' depending on time. The solution presented below is inspired
from the results of Ref. $\left[ 7\right] $ and we show that our
cosmological constant $\Lambda $ is related with the constant $H_{0}$ in
that work and which is expected to be Planck scale.

The calculations in this paper, especially in Sect. 3, were performed using
an analytical program written by us in the package GRTensor II running on
the Maple V platform. This program allows to calculate the components $%
F_{\mu \nu }^{a}$ and $F_{\mu \nu }^{ab}$ of the strength tensor field $%
F_{\mu \nu }$, the expressions of the quantities $F_{\mu }^{a}$ defined in
Eq. (15), and the invariants $F$, $I_{1}$, $I_{2}$. We calculated also the
integrand in the action $S_{g}$ and obtained the field Eqs. (25)-(27).

It is important to emphasize that in our gauge model of gravitation we do
not use a metric, but only the gauge fields (potentials) $e_{\mu }^{a}\left(
x\right) $ and $\omega _{\mu }^{ab}\left( x\right) $ of the gravitational
field. In program we used the notations: einv\{a \symbol{94}mu\}= $\overset{%
\_}{e}_{a}^{\mu }$ for the inverse of $e_{\mu }^{a}$, and $de=\det \left(
e_{\mu }^{a}\right) $ for the determinant of $e_{\mu }^{a}$. The symbols for
other quantities are introduced analogously.

Below, we list down the part of program which allows to define and to
calculate the quantities needed in obtaining of Eqs. (25)-(27).

Program '' DS GAUGE THEORY.MWS''

\TEXTsymbol{>} restart:

\TEXTsymbol{>} grtw( ):

\TEXTsymbol{>} grload(minkowski, `d:/maple/sferice.mpl`);

\TEXTsymbol{>} grdef(`ev\{\symbol{94}a mu\}`); grcalc(ev(up,dn));

\TEXTsymbol{>}grdef(`eta1\{(a b)\}`); grcalc(eta1(dn,dn))

\TEXTsymbol{>} grdef(`omega\{[\symbol{94}a \symbol{94}b] mu\}`);
grcalc(omega(up,up,dn));

\TEXTsymbol{>} grdef(`Famn\{\symbol{94}a mu nu\} := ev\{\symbol{94}a nu,mu\}
- ev\{\symbol{94}a mu,nu\}

+ omega\{\symbol{94}a \symbol{94}b mu\}*ev\{\symbol{94}c nu\}*eta1\{b c\}

- omega\{\symbol{94}a \symbol{94}b nu\}*ev\{\symbol{94}c mu\}*eta1\{b c\}`);

\TEXTsymbol{>} grcalc(Famn(up,dn,dn));

\TEXTsymbol{>}grdef(`Fabmn\{\symbol{94}a \symbol{94}b mu nu\} := omega\{%
\symbol{94}a \symbol{94}b nu, mu\}-

omega\{\symbol{94}a \symbol{94}b mu, nu\}+ (omega\{\symbol{94}a \symbol{94}c
mu\}*omega\{\symbol{94}d \symbol{94}b nu\} -

omega\{\symbol{94}a \symbol{94}c nu\}*omega\{\symbol{94}d \symbol{94}b
mu\})*eta1\{c d\}-

\ 4*lambda\symbol{94}2*( ev\{\symbol{94}a mu\}*ev\{\symbol{94}b nu\} - ev\{%
\symbol{94}b mu\}*ev\{\symbol{94}a nu\})`);

\TEXTsymbol{>} grcalc(Rabmn(up,up,dn,dn));

\TEXTsymbol{>} grdef(`R:=Rabmn\{\symbol{94}a \symbol{94}b mu nu\}*einv\{a 
\symbol{94}mu\}*einv\{b \symbol{94}nu\}`);

\TEXTsymbol{>} grcalc(R);

\TEXTsymbol{>} grdef(`F\{\symbol{94}a mu\}:=Rabmn\{\symbol{94}a \symbol{94}b
mu nu\}*einv\{b \symbol{94}nu\}`);

\TEXTsymbol{>} grcalc(F(up,dn));

\TEXTsymbol{>} grdef(`I2:=4*F\{\symbol{94}a mu\}*Finv\{a \symbol{94}mu\}-(R)%
\symbol{94}2`); grcalc(I2);

\TEXTsymbol{>}grdef(`I1:=R-sqrt(3)*sqrt(I2)`); \TEXTsymbol{>} grcalc(I1);

\TEXTsymbol{>} grdef(`de`); grcalc(de);

\TEXTsymbol{>} grdef(`Lg:=(R+phi1(t)*I1-phi2(t)*sqrt(I2)+V1(phi1)+

V2(phi2))*de`); grcalc(Lg); grdisplay(\_);

\section{Example of solution without singularities}

The solution of Eqs. (25)-(27) includes a dependence on the cosmological
constant $\Lambda $. We suppose that the Lagrange-multiplier function $%
\varphi _{1}\left( t\right) $ is absent, and consider the case when $k=0$.
Then, denoting $\varphi _{2}\left( t\right) =\varphi \left( t\right) $ and $%
V_{2}\left( \varphi _{2}\right) =V\left( \varphi \right) $, the Eqs.
(25)-(27) become: 
\begin{eqnarray}
H^{\prime } &=&-\frac{1}{2\sqrt{3}}\frac{dV}{d\varphi },  \TCItag{29} \\
\varphi ^{\prime } &=&-3H\varphi +\sqrt{3}H-\frac{1}{2\sqrt{3}H}V-\frac{%
2\Lambda }{\sqrt{3}H}.  \notag
\end{eqnarray}
We consider the potential $V\left( \varphi \right) $ of the simple form: 
\begin{equation}
V\left( \varphi \right) =2\sqrt{3}\lambda ^{2}\left( \frac{\varphi ^{2}}{%
1+\varphi ^{2}}+\frac{24}{\sqrt{3}}\right) ,  \tag{30}
\end{equation}
where $\lambda $ is the real parameter that determines the cosmological
constant $\Lambda $. This parameter coincides with the constant $H_{0}$ in
Ref. $\left[ 7\right] $ that has been interpreted as a Planck scale of the
model. Therefore, in our example the Planck scale is related to the
cosmological constant $\Lambda $. For small values of $H$ and $\varphi $,
the Eqs. (29) can be written as: 
\begin{eqnarray}
H^{\prime }\text{ } &\simeq &-2\lambda ^{2}\varphi ,  \TCItag{31} \\
\varphi ^{\prime }\left( t\right) &\simeq &\frac{\sqrt{3}H^{2}-\lambda
^{2}\varphi ^{2}}{H}.  \notag
\end{eqnarray}

These equations have the periodic solution 
\begin{equation}
\varphi \left( t\right) =\varphi _{0}\sin \left( \omega t\right) ,\text{ \ \ 
}H\left( t\right) =\frac{\omega \varphi _{0}}{2\sqrt{3}}\left[ \cos \left(
\omega t\right) -1\right] ,  \tag{32}
\end{equation}
where $\varphi _{0}$ is an integration constant and $\omega =2\times
3^{1/4}\lambda $ is the frequency of oscillation of the corresponding
gravitational field described by the gauge potentials $e_{\mu }^{a}\left(
x\right) $ and $\omega _{\mu }^{ab}\left( x\right) $. This solution has no
singularities and it is valid if the cosmological constant is negative ($%
\Lambda <0$). The case with positive cosmological constant ($\Lambda >0$)
can be studied choosing the anti-de-Sitter group as gauge group. But, the
deformation parameter $\lambda $ will be then pure imaginary. We emphasize
that there are possible also periodic solutions if we suppose a
time-dependent cosmological ''constant''. In particular, we can consider a
cosmological ''constant'' which is itself a periodic function on time. It
will be also of interest to apply the previous method in obtaining
non-singular solutions of the gauge theories with internal groups of
symmetry.

\section{Concluding remarks}

We developed a de-Sitter gauge theory of gravitation on a spherical
symmetric Minkowski space-time as base manifold. This theory allows a
complementary description of the gravitational effects in which the
mathematical structure of the underlying space-time is not affected by
physical events.$^{9}$ Only the gauge potentials $e_{\mu }^{a}\left(
x\right) $ and $\omega _{\mu }^{ab}\left( x\right) $ of the gravitational
field change as functions of coordinates. This is important when we consider
a quantum gauge theory of gravitation.

In order to obtain solutions without singularities, we imposed constraints
on some invariants of the gauge model we considered. The solutions in this
paper are time-periodic and correspond to a fixed (negative) cosmological
constant $\Lambda $ whose value is related to the Planck scale and that
determines the frequency of the corresponding gravitational field.

\bigskip

References

1. M. Blagojevi\'{c}, Three lectures on Poincar\'{e} gauge theory,

arXiv:gr-qc/0302040, v1 11 Feb 2003

2. G. Zet, V. Manta, S. Babeti, Int.J.Mod.Phys. C14, 41 (2003)

3. F. Gronwald, Int.J.Mod.Phys. D6, 263(1997)

4. S.W. MacDowell, F. Mansouri, Phys.Rev.Lett. 38,739 (1977)

5. M. Calcada, J.G. Pereira, Int.J.Theor.Phys. 41, 729 (2002)

6. G. Zet, Schwarzschild solution on a space-time with torsion,
arXiv:gr-qc/0308078

7. R. Brandenberger, M. Mukhanov, A. Sornborger, Phys.Rev. D48, 1629 (1993)

8. P. Freund, Supersymmetry, Cambridge Univ. Press, N.Y. 1983

9. C. Wiesendanger, Classical and Quantum Gravity 13, 681 (1996)

\end{document}